\documentclass[aps,reprint,longbibliography,showkeys,superscriptaddress]{revtex4-2}
\usepackage{bm,graphicx,tabularx,array,booktabs,dcolumn,xcolor,microtype,multirow,amscd,amsmath,amssymb,amsfonts,physics,tikz,wrapfig,enumitem,float,upgreek}

\usepackage[version=4]{mhchem}
\usepackage[utf8]{inputenc}
\usepackage[T1]{fontenc}
\usepackage[normalem]{ulem}
\usepackage{times}

\usepackage{}
\usepackage[]{graphicx}
\usepackage{setspace}
\usepackage{siunitx}
\usepackage{fixltx2e}
\DeclareSIUnit\hartree{E\textsubscript{h}}

\usepackage{hyperref}
\hypersetup{
    colorlinks=true,
    linkcolor=blue,
    filecolor=blue,      
    urlcolor=blue,	
    citecolor=blue
}

\usepackage{titlesec}
\titleformat{\section}{\bfseries}{}{0em}{}
\titleformat{\subsection}{\it\bfseries}{}{0em}{}
\titleformat{\subsubsection}{\it}{}{0em}{}
\titlespacing\section{0pt}{10pt plus 2pt minus 2pt}{2pt plus 2pt minus 2pt}
\titlespacing\subsection{0pt}{10pt plus 2pt minus 2pt}{2pt plus 2pt minus 2pt}
\titlespacing\subsubsection{0pt}{10pt plus 2pt minus 2pt}{2pt plus 2pt minus 2pt}
\usetikzlibrary{quantikz2}

\usepackage{caption} % Allow width of caption to be controlled
\usepackage{subcaption}
\usepackage{cleveref}
\renewcommand\cite[1]{(\citenum{#1})}

\newcommand{\hH}{\hat{H}}
\newcommand{\hHe}{\hat{H}_{\text{e}}}

\newcommand{\cT}{\mathcal{T}}

\newcommand{\ibmtorino}{\texttt{ibmq\_torino}}

\newcommand{\wmax}{\omega_{\text{max}}}
\newcommand{\lmax}{\lambda_{\text{max}}}

\definecolor{hughgreen}{RGB}{100, 0, 140}

\allowdisplaybreaks

\newcommand{\UCAM}{Yusuf Hamied Department of Chemistry, University of Cambridge, Lensfield Road, Cambridge, CB2 1EW, U.K.}
\newcommand{\UCL}{Department of Chemistry, University College London, London, WC1H 0AJ, U.K.}

\begin{document}	

%\title{Excited adiabatic state preparation on a quantum computer using electron-photon coupling}
\title{Excited state preparation on a quantum computer through adiabatic light-matter coupling}

\date{\today}
\author{Hugh G.~A.~\surname{Burton}}
\email{h.burton@ucl.ac.uk}
\affiliation{\UCL}
%\affiliation{\UCAM}
\author{Maria-Andreea~\surname{Filip}}
\affiliation{\UCAM}

\begin{abstract}
\noindent\normalsize
Quantum computing has the potential to transform simulations of quantum many-body problems 
at the heart of electronic structure theory. 
Efficient quantum algorithms to compute the eigenstates of fermionic Hamiltonians, such as quantum phase estimation, rely critically on high-accuracy initial state preparation.
While several state preparation algorithms have been proposed for fermionic ground states, 
the preparation of excited states remains a major challenge, limiting the applicability of quantum algorithms
to photochemistry and photophysics.
In this contribution, we describe a physically motivated adiabatic state preparation technique for low-lying bright excited states 
using the explicit coupling between electrons and photons. 
Our approach systematically converges to the first bright excited state and 
can target different symmetry sectors by changing the photon polarization.
We demonstrate the preparation of high-fidelity excited states for the Hubbard model and methylene 
molecule across a range of correlation regimes,
and perform a successful hardware implementation for a model Hamiltonian. 
\end{abstract}

\maketitle
\raggedbottom
\onecolumngrid

\vspace{-2em}
%%%%%%%%%%%%%%%%%%%%%%%%
\section{Introduction}
\label{sec:intro}
%%%%%%%%%%%%%%%%%%%%%%%%

% KEY IDEA IS EXCITED STATES. THEY'RE HARD, BUT VERY IMPORTANT
Computing excited-state electronic wavefunctions %{\color{red} (MAF: I don't really like ``solving wavefunctions")} 
is central to theoretically understanding interactions between light and matter, 
enabling predictions of optical material properties, 
absorption and emission spectra, and photochemical reaction pathways. 
However, excited states are more challenging to solve
than ground states because the variational principle generally cannot be used to find the best wavefunction approximation.
We currently rely on multiconfigurational theory,
which scales exponentially with system size,
or linear response methods that can only describe single-electron excitations and are only accurate
if a good ground-state approximation can be found.\cite{Gonzalez2011,Lischka2018} 
Accurate excited-state predictions for molecules and materials with strong electron correlation or 
multi-electron excitations remain a major theoretical challenge.

% QUANTUM ALGORITHMS FOR QUANTUM COMPUTATION
Electronic structure theory is poised to be one of the first applications of quantum computers, benefiting
from the fact that a linear number of qubits can encode the exponentially scaling 
electronic Hilbert space.\cite{Cao2019,McArdle2020,Motta2022}
Hybrid algorithms, such as the variational quantum eigensolver (VQE),\cite{Peruzzo2014,Cerezo2021a}
can generate highly accurate wavefunctions, but require careful ansatz
definition and often suffer from optimisation difficulties.\cite{McClean2018,Wang2021,Cerezo2021}
The success of many other near-term and fault-tolerant quantum algorithms depends on the overlap between an initial trial state $\ket{\Psi_\text{T}}$, prepared on the quantum device, 
and the true eigenstate $\ket{\Psi_I}$ of interest.\cite{Lee2023}
For example, the probability of measuring the ground-state energy $E_0$ 
using quantum phase estimation scales as $p(E_0)\propto\abs{\braket{\Psi_\text{T}}{\Psi_0}}^2$.\cite{AspuruGuzik2005}
To address this challenge, several high-accuracy state-preparation methods have been 
established for ground states, 
including adiabatic state preparation (ASP),\cite{AspuruGuzik2005,Albash2018} 
VQE,
and circuits to encode high-accuracy wavefunctions that have been precomputed 
on a conventional computer.\cite{Formichev2024}
In contrast, algorithms to prepare excited states are much less explored and have focused
almost exclusively on hybrid VQE 
methods.\cite{Nakanishi2019,Parrish2019,Yalouz2021,Grimsley2025,McClean2016,Santagati2018,CadiTazi2024,Lee2019,Higgott2019,Chan2021,Ibe2022,Shirai2022}

% VISION THAT WE IMPLEMENT HERE
In this work, we introduce an excited state preparation technique that uses quantum adiabatic state preparation to
directly convert between ground and excited states by leveraging the physical coupling between photons and electrons.
Our approach starts with the electronic ground state $\ket{\Psi_0}$ 
and a singly-occupied photon state  $\ket{1}$
encoded on a quantum register as the multicomponent product state $\ket{\Psi_0}\otimes\ket{1}$.
Starting with the initial photon frequency $\omega=0$, 
the state is adiabatically evolved while increasing $\omega$ and 
explicitly coupling the electronic and photonic modes through the dipole operator.
By following a suitable pathway, the combined state can be continuously evolved into the first bright excited electronic state $\ket{\Psi_{\text{es}}}$ coupled to the  photon vacuum state, i.e.\  $\ket{\Psi_\text{es}}\otimes\ket{0}$.
This approach artificially controls the physical coupling between electrons and photons in a photoexcitation,
allowing the ground state to be adiabatically evolved into the lowest 
optically accessible excited state without any prior knowledge of the excited-state wavefunction.

% HOW THIS FITS IN EXISTING EXCITED-STATE METHODS....
Current excited-state quantum algorithms are primarily based on
VQE or diagonalization-based methods. 
Excited-state VQE algorithms employ modified objective functions, such as 
the ensemble energy of multiple low-lying states\cite{Nakanishi2019,Parrish2019,Yalouz2021,Grimsley2025} 
and the folded-spectrum approach,\cite{McClean2016,Santagati2018,CadiTazi2024}
or enforce orthogonality to the ground state and 
lower-lying excited states.\cite{Lee2019,Higgott2019,Chan2021,Ibe2022,Shirai2022}
These methods all require a suitable variational \textit{ansatz} and face practical challenges 
such as difficult convergence, high measurement costs, and deep quantum circuits.
Furthermore, the multitude of stationary points in variational algorithms\cite{Boy2025} 
may hinder excited-state assignment as the variational principle no longer indicates
the most physical approximation. 
Alternatively, quantum equation of motion and linear response theory\cite{Colless2018,Ollitrault2020,Urbanek2020,Asthana2023,Kumar2023,Jensen2024,Reinholdt2024,Gandon2024,Ziems2024} can be used
to compute excited-state energies and properties from a VQE ground-state wavefunction prepared on the quantum device. 
However, this method only provides access to energies and is not suitable for excited state preparation.

Another approach is to use quantum subspace diagonalization techniques, where the quantum device is used 
to compute Hamiltonian and overlap matrix elements for a set of basis states and the corresponding eigenstates are computed on a classical device.\cite{Motta2024}
Variants of this method differ in the choice of basis states.
The quantum subspace expansion (QSE) uses basis states constructed using excitations from a ground-state VQE solution,\cite{McClean2017} 
mirroring traditional multireference configuration interaction techniques.\cite{Lischka2018}
Quantum Krylov (QKS) methods build the basis as a Krylov expansion using real or imaginary time evolution.\cite{Stair2020,Motta2020,Cortes2022,Klymko2022,Kirby2023,Oumarou2025,Oliveira2025}
However, both QSE and QKE require the solution of a generalized eigenvalue problem, which can become unstable in the presence of hardware noise.\cite{Lee2025,Cortes2022}
Furthermore, preparing the resulting excited states on the quantum device requires a linear combination of 
basis states to be encoded using many ancilla qubits and controlled operations.\cite{Formichev2024}

Diagonalization-based approaches also include quantum selected configuration interaction (QSCI) and 
sample-based quantum diagonalization (SQD).\cite{Kanno2023,Reinholdt2025,Mikkelsen2025,RobledoMoreno2025,Barison2025}
In these methods, the most important configurations are sampled from a trial state prepared on the quantum device
and then used as a basis for truncated configuration interaction on classical hardware.
In contrast to quantum subspace diagonalization, the quantum device is only used to sample (orthogonal) computational basis state,
while the Hamiltonian matrix elements are evaluated using the Slater--Condon rules.
This feature makes QSCI/SQD well-suited to near-term devices with high noise levels.
While the sampling yields important ground-state configurations, excited states can also be computed by augmenting the basis by applying excitations to the sampled configurations.\cite{Barison2025}
However, QSCI/SQD techniques scale exponentially with respect to system size due to the final configuration interaction step,
and rely heavily on finding a good trial distribution to avoid inefficient sampling.\cite{Reinholdt2025}
Combined with the need to prepare the final linear combination of configurations on quantum hardware, these features mean that excited state preparation using QSCI/SQD would be challenging.

% A BIT ABOUT THE ADIABATIC STATE PREPARATION, WHY IT MIGHT BE BETTER
As an alternative to hybrid quantum algorithms, adiabatic state preparation (ASP) is a well-established method for preparing fermionic 
ground states,\cite{AspuruGuzik2005,Veis2014,Albash2018,Kremenetski2021,Formichev2024,Granet2025,MartiDafcik2025} 
although it is much less explored than methods based on VQE and QSD.
After initialising the quantum state as an eigenstate of a reference Hamiltonian $\hH_\text{init}$, 
for which the ground state can be easily prepared, the system undergoes time evolution
as the Hamiltonian is slowly changed to match the target system $\hH_\text{final}$.
If the time evolution is sufficiently slow, then the adiabatic theorem ensures that 
the system remains in an eigenstate at all times, becoming the target eigenstate at the end of the 
evolution.\cite{Born1928}
The success of adiabatic time evolution is controlled by the minimum energy gap along the 
pathway, with a smaller gap requiring longer time evolution.\cite{Albash2018}
Compared to VQE, adiabatic methods have more robust theoretical guarantees of success and 
rely on fewer heuristics, such as the definition of a variational ansatz.
Previous investigations for quantum chemistry have focussed on ground state preparation, 
with $\hH_\text{init}$ defined using a one-body Fock operator\cite{Veis2014} or from the diagonal of the 
many-body Hamiltonian.\cite{AspuruGuzik2005,Kremenetski2021}
However, to the best of our knowledge, no ASP algorithms have yet been proposed
for excited state preparation.

%\it Adiabatic state preparation (ASP) has previously been proposed as a method to prepare ground states 
%on quantum device by taking advantage of the adiabatic theorem.\cite{Albash2018,Kremenetski2021}
%There is not much literature on preparing excited states via ASP and where it exists it generally
%employs conventional ASP starting from some excited
%state of the initial Hamiltonian, under the assumption that if the adiabatic condition is satisfied, this
%state will be preserved. Yarloo \textit{et al}\cite{Yarloo2024} propose using ASP to obtain highly excited scar states in a
%matrix product state model. While these are close in energy to thermal excited states, they have much lower
%entanglement entropy, which seems to limit transfer to other states during ASP.
%Hwang \textit{et al}\cite{Hwang2024} propose using ASP together with the CoVar algorithm (which finds
%eigenstates by minimising a large number of Hamiltonian covariances) to obtain both ground and excited states.
%Generally, the state found is not well controlled.

% RESULTS AND FINDINGS
% HGAB: Here, just trying to pull together the key "achievements" of the paper.
Here, we describe the Excited Adiabatic State Preparation (EXASP) algorithm to 
prepare the lowest-energy bright excited state on a quantum computer.
We define an adiabatic evolution pathway to construct high-fidelity
excited-state wavefunctions on a quantum register and show how the fidelity
%{\color{red} (MAF: fidelity is enhanced, at the cost of non-unity success probability)}
can be further enhanced using post-measurement selection.
Our approach requires minimal prior information about the target excited-state wavefunction 
and is free from any variational ansatz or diagonalization.
Furthermore, we show how specific molecular excited states can be targeted using
symmetry selection rules.
Our approach can prepare high-accuracy excited states for both weakly and strongly 
correlated Hamiltonians, and provides a fault-tolerant
route towards quantum initial state preparation for theoretical photochemistry.

%%%%%%%%%%%%%%%%%%%%%%%%
\section{Results}
\label{sec:results}
%%%%%%%%%%%%%%%%%%%%%%%%
%%%%%%%%%%%%%%%%%%%%%%%%
\subsection{Excited adiabatic state preparation}
\label{subsec:EXASP}
%%%%%%%%%%%%%%%%%%%%%%%%

% Here, we introduce the adiabatic state preparation in more detail. 
% Then, we show how this can be combined with a multicomponent Hamiltonian to predict an excited state.
% Figure: Schematic showing avoided crossing for different lambda values

% OVERVIEW OF GROUND-STATE ASP

The adiabatic theorem states that a system will remain in an instantaneous eigenstate if 
changes to the Hamiltonian occur sufficiently slowly and there is a non-zero gap between the corresponding 
eigenvalue and the rest of the spectrum.\cite{Born1928}
Quantum adiabatic state preparation %{\color{red} (MAF: Same comment about adiabatic quantum computing as before)}
 exploits this theorem by defining 
an initial Hamiltonian $\hH_\text{init}$, with known eigenstates
that can be easily prepared on the quantum device,
and a target Hamiltonian $\hH_\text{final}$ with unknown eigenstates.\cite{Albash2018}
A system prepared initially in an eigenstate of $\hH_\text{init}$ is then evolved for a 
total time $T$ through the time-dependent Schr\"{o}dinger equation with $\hbar = 1$ as
\begin{equation}
\frac{\mathrm{i}}{T} \frac{\partial }{\partial s}\ket{\Psi(s)} = \hat{H}(s) \ket{\Psi(s)},
\label{eq:tdse}
\end{equation}
where $s = t/T \in [0,1]$ is a dimensionless path coordinate and $\hat{H}(s)$
is defined such that $\hat{H}(0) = \hat{H}_\text{init}$ and $\hat{H}(1) = \hat{H}_\text{final}$.
Note that if $\hbar = 1$, then the units of $T$ are the inverse of the natural energy unit for the corresponding Hamiltonian.
As long as $\hat{H}(s)$ evolves sufficiently slowly along the path, and remains gapless 
for the eigenstate of interest, then the system will remain in an exact eigenstate for 
$T \rightarrow \infty$.
The solution to Eq.~\eqref{eq:tdse}  is obtained using the time-evolution operator
\begin{equation}
\hat{U}(s) = \cT \qty[\exp(\int_0^s - \mathrm{i} T \hat{H}(s')\, \dd s')],
\label{eq:timeOp}
\end{equation}
where $\cT$ is the time-ordering operator, giving the time-evolved wavefunction
%\begin{equation}
$\ket{\Psi(s)} = \hat{U}(s) \ket{\Psi(0)}$.
%\end{equation}
In practice, the time evolution is discretized into $N$ steps using a first-order Trotter approximation
\begin{equation}
\hat{U}(s) \approx \exp(-\mathrm{i}\,\delta s\, T \hat{H}(s_{N-1})) \cdots  \exp(-\mathrm{i}\,\delta s \,T \hat{H}(s_0))
\label{eq:Us}
\end{equation}
where  $\delta s\,T = \delta T$ is the time step, $\delta s = N^{-1}$, and $s_k = k/N$.
The Trotter approximation becomes exact in the limit $\delta T \rightarrow 0$, while exact adiabatic 
evolution is achieved for $T \rightarrow \infty$.
The number of steps $N$ required to achieve satisfactory accuracy controls the quantum 
circuit depth and the computational cost of ASP.\cite{Albash2018,Veis2014,Kremenetski2021}
%Bounds have been derived that show $T$ is inversely proportional to the energy gap between the ground and the first excited state along the adiabatic path.

%%%%%%%%%%%%%%%%%%%%%%%%%%%%%%%%%%%%%%%%%%%%%%%%%%%%
%\begin{figure}[htb]
%	\includegraphics[width=\linewidth]{../two_level.pdf}
%	\caption{\small Eigenvalues for the two-level model coupled to a signle photon mode [Eq.~\eqref{eq:ham_twolevel}].
%		Dashed lines indicate the eigenstates without any electron-photon coupling. 
%		States with energy proportional to $\omega$ correspond to photonic modes. }
%	\label{fig:twolevel} 
%\end{figure}
%%%%%%%%%%%%%%%%%%%%%%%%%%%%%%%%%%%%%%%%%%%%%%%%%%%%

% INTRODUCE 
To define an adiabatic pathway that evolves a ground state into an excited state, we consider a multicomponent
system where the electrons are explicitly coupled to a single two-level photon mode with frequency $\omega$.
This combined system can be described in the long-wavelength dipole limit and the length gauge using
the Pauli--Fierz Hamiltonian\cite{SpohnBook}
\begin{equation}
\hH = \hHe + \omega\, \hat{b}^\dagger \hat{b} 
- \lambda \sqrt{\frac{\omega}{2}} (\bm{e} \cdot \hat{\bm{\mu}} ) (\hat{b}^\dagger + \hat{b}) + \frac{\lambda^2}{2} (\bm{e}\cdot\hat{\bm{\mu}})^2,
\label{eq:PFham}
\end{equation}
where the purely electronic part is defined as usual\cite{HelgakerBook}
\begin{equation}
	\hHe = V_{\text{ext}} + \sum_{pq} \hat{a}^\dagger_{p} \hat{a}_{q} h_{pq} 
	+ \frac{1}{2} \sum_{pqrs} \langle pq|rs \rangle \hat{a}^\dagger_{p} \hat{a}^\dagger_{q} \hat{a}_{s}  \hat{a}_{r}.
\end{equation}
Here, $\hat{a}_p$ and $\hat{b}$ are the second-quantized field operators for the electron and photon modes respectively, 
$\lambda$ controls the electron\nobreakdash-photon coupling strength,
$\bm{e}$ is the photon polarization vector, and $\hat{\bm{\mu}}$ is the electronic dipole operator.
The electron\nobreakdash-photon interaction can excite (de-excite) electrons through the dipole operator while simultaneously
annihilating (creating) a photon.
The last term in Eq.~\eqref{eq:PFham} corresponds to the dipole self-energy and is required to ensure that the 
Hamiltonian is bounded from below in the general case.\cite{Rokaj2018}
This Hamiltonian derives from the coupling of a molecule to an optical cavity, which provides a physical 
realization for a quantized photon field with a single mode.\cite{Rokaj2018,Mordovina2020,Haugland2020,Ruggenthaler2023}
In this work, the photon mode is a purely fictitious interaction used to control the adiabatic state evolution, and thus 
we are free to vary $\omega$, $\lambda$, and $\bm{e}$ without physical constraints. 
Since $\omega$ is of similar magnitude to the electronic excitation energies, the long-wavelength limit assumed in Eq.~\eqref{eq:PFham} is valid.
We employ the Pauli--Fierz form of the QED Hamiltonian in the length gauge because it is widely used in polariton chemistry,\cite{Haugland2020,Mordovina2020,Ruggenthaler2014,Ruggenthaler2023,Flick2017,Rokaj2018}
and it provides a strictly real Hamiltonian where the pure electronic term is identical to the standard molecular Hamiltonian.\cite{Taylor2025}
%and shows better convergence properties with
%respect to the electronic basis set than the velocity gauge.\cite{DeBernardis2018,Li2020,Taylor2020}}

\begin{figure*}[htb!]
	\includegraphics[width=\linewidth]{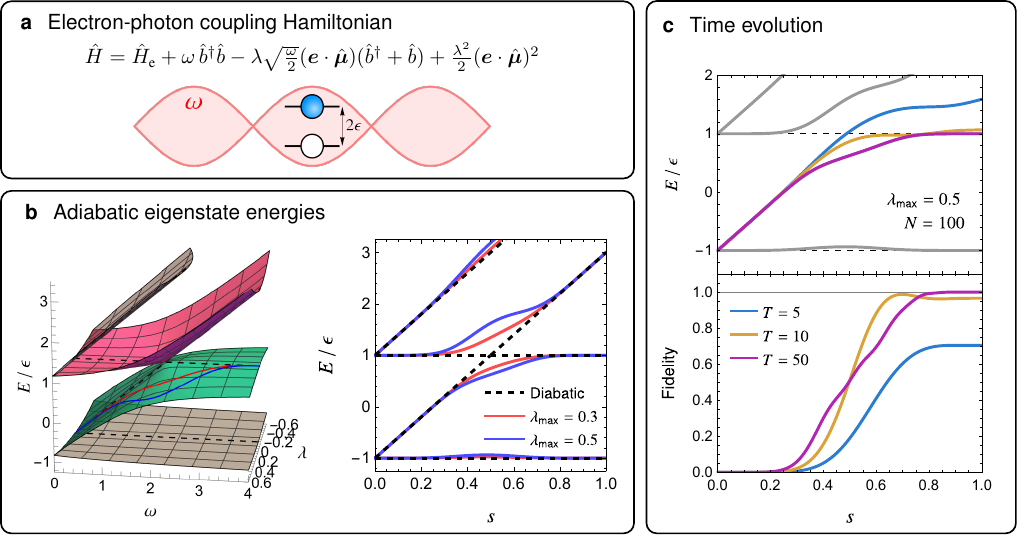}
	\caption{\textbf{Explicit electron-photon coupling enables adiabatic preparation of an excited-state wavefunction.}
		(\textbf{\textsf{a}}) The electron-photon coupling Hamiltonian 
		for a two-level system is analogous to the Jaynes--Cummings model.
		(\textbf{\textsf{b}}) The adiabatic states form a
		conical intersection in the $(\omega,\lambda)$ plane (left). 
		Following a suitable parametrized pathway [Eq.~\eqref{eq:path}] provides a connection between 
		the electronic ground and excited state (right), with $\lmax$ controlling the strength of the 
		avoided crossing.
		(\textbf{\textsf{c}}) Adiabatic time evolution along this parametrized pathway $\hH(s)$ 
		enables the preparation of the excited-state wavefunction, with a success probability 
		that depends on the total evolution time $T$.
	}
	\label{fig:twolevel}
\end{figure*}

We describe how the electron\nobreakdash-photon coupling can be used to adiabatically connect ground and excited states using 
a two-level electronic system with eigenenergies $\pm \epsilon$ and the dipole coupling $\mel{\Psi_0}{\bm{e}\cdot\hat{\bm{\mu}}}{\Psi_1}= \mu$, 
analogous to the Jaynes--Cummings model\cite{Jaynes1963} (Fig.~\ref{fig:twolevel}\textcolor{blue}{a}).
The diabatic basis for this system is $\{\ket{0; 0}, \ket{0; 1}, \ket{1; 0},\ket{1; 1} \}$, 
where $\ket{I;n} = \ket{I}\otimes\ket{n}$ denotes the tensor product of the electronic eigenstate $\ket{I} \equiv \ket{\Psi_I}$ and a
photon mode $\ket{n}$ with occupation $n$. 
In this two-level Jaynes--Cummings model, the dipole self-energy  $\frac{\lambda^2}{2} (\bm{e}\cdot\hat{\bm{\mu}})^2$ in Eq.~\eqref{eq:PFham} provides a 
uniform energy shift and can therefore be ignored.
The Hamiltonian in this basis is then
\begin{equation}
\hH = 
\begin{pmatrix}
-\epsilon & 0 & 0 & \lambda \mu \sqrt{\omega/2} \\ 
0  & -\epsilon + \omega & \lambda \mu \sqrt{\omega/2} & 0  \\ 
0  & \lambda \mu \sqrt{\omega/2} & \epsilon  & 0  \\ 
 \lambda \mu \sqrt{\omega/2} & 0 & 0 & \epsilon + \omega\\ 
\end{pmatrix}.
\label{eq:twolevelH}
\end{equation}
The lowest eigenstate of Eq.~\eqref{eq:twolevelH} remains almost exclusively in the diabatic state $\ket{0;0}$ for all positive $\omega$.
In contrast, in the $(\omega,\lambda)$ plane,  the  second eigenstate evolves continuously from the electronic ground state 
$\ket{0;1}$ for $\omega \rightarrow 0$ to the electronic excited state $\ket{1;0}$ for $\omega \gg 2\epsilon$, mirroring the annihilation of 
a photon during an electronic excitation (Fig.~\ref{fig:twolevel}\textcolor{blue}{b}).
A conical intersection between the diabatic states $\ket{0;1}$ and $\ket{1;0}$ occurs at $(\omega,\lambda) = (2\epsilon,0)$, where the 
electronic excitation and the photon frequency are at resonance but the coupling strength is zero.
Following a suitable pathway that connects the diabatic states $\ket{0;1}$ and $\ket{1;0}$  in the $(\omega,\lambda)$ plane, 
while avoiding the conical intersection, allows the excited state to be adiabatically prepared from the ground state starting point  (Fig.~\ref{fig:twolevel}\textcolor{blue}{b}; left).
The definition of a suitable adiabatic pathway, and the implementation on a quantum device, are described in the section \textbf{\textit{Definition of the adiabatic pathway}} below.

%%%%%%%%%%%%%%%%%%%%%%%%%%%%%
%%%%%%%%%%%%%%%%%%%%%%%%%%%%%

% GENERALITY AND ADVANTAGES
This approach is applicable to any electronic Hamiltonian.
Assuming that the ground state can be prepared on the quantum device and a suitable pathway can be defined, 
the first optically active (bright) excited state can be prepared without any prior knowledge of the excited-state wavefunction.
Furthermore, since the fictitious photon mode is encoded as an additional qubit, post-selection 
on the photon qubit can be used to project the 
final state into the photon vacuum state, improving the fidelity of the final excited 
state obtained by adiabatic evolution.
While this post-selection increases the success probability of a subsequent
QPE calculation ($p_\text{QPE}$), the overall success probability $p_\text{tot} = p_\text{post}p_\text{QPE}$
also depends on the probability of successful post-selection $p_\text{post}$ and is the same as applying QPE directly to the state prior to post-selection.
However, the advantage of post selection is that it allows unsuccessful runs to be filtered out before expensive quantum
processing of the adiabatically prepared state, reducing the overall quantum resource cost.

%%%%%%%%%%%%%%%%%%%%%%%%
\subsection{Definition of the adiabatic pathway}
%%%%%%%%%%%%%%%%%%%%%%%%

The adiabatic evolution $\hH(s)$ for $s\in[0,1]$ is constructed by defining parametric functions $\omega(s)$ and $\lambda(s)$.
The functional forms of $\omega(s)$ and $\lambda(s)$ are constrained by the conditions $\omega(0)=0$ and $\lambda(0)=\lambda(1)= 0$,
and should be defined such that an excited state is obtained for $s=1$.
The condition $\lambda(0)=\lambda(1)= 0$ ensures that the electron-photon coupling vanishes in the initial and final Hamiltonians.
Furthermore, successful adiabatic state preparation requires that the total evolution time $T$
satisfies the condition\cite{Albash2018,Amin2009}
\begin{equation}
T \gg \max_{K\neq J} \max_{s\in[0,1]}\frac{\abs*{\mel{\Psi_K(s)}{\partial_s \hH(s)}{\Psi_J(s)}}}{\abs{E_K(s) - E_J(s)}^2},
\label{eq:Tcondition}
\end{equation}
where $\{ \ket{\Psi_K(s)}, E_K(s)\}$ are the instantaneous eigenpairs of $\hH(s)$ and $\ket{\Psi_J(s)}$ is the eigenstate involved in the adiabatic evolution with energy $E_J(S)$. 
This condition derives from the probability that a non-adiabatic transition occurs between two eigenstates of $\hH(s)$ during the time evolution, 
causing the system to finish in the wrong eigenstate at the end of the time evolution.
The usual interpretation of this condition is that adiabatic evolution requires the 
target eigenstate to be non-degenerate along the full pathway.
A more precise definition is that the full fraction Eq.~\eqref{eq:Tcondition} must be near-zero
everywhere, meaning that the denominator must be large or the numerator must vanish.

% NEAR DEGENERACIES
There are three cases where $\abs{E_K(s) - E_J(s)}$ can become small during 
the EXASP evolution.
(i) At the avoided crossing, where the system transitions from the diabatic ground state with a 
photon to the excited state without a photon.
In this case, the gap may be artificially widened by increasing the electron-photon coupling $\lambda$ to avoid 
non-adiabatic transitions (Fig.~\ref{fig:twolevel}\textcolor{blue}{b}; right).
(ii) If two eigenstates become exactly degenerate and their energies intersect along the pathway. 
From the non-crossing theorem with one parameter $s$, this can only occur if the eigenstates do not couple through $\hH(s)$ due to 
symmetry. 
Since $\partial_s \hH(s)$ shares the same symmetries as $\hH(s)$, then
if $\mel{\Psi_K(s)}{\hH(s)}{\Psi_J(s)} $ vanishes due to symmetry, so will $\mel{\Psi_K(s)}{\partial_s \hH(s)}{\Psi_J(s)}$.
Therefore, the numerator in Eq.~\eqref{eq:Tcondition} will vanish for all $s$ and 
no transition can occur between the two states during the adiabatic evolution.
%and thus the adiabatic evolution is unaffected.
%that do not couple through $\hH(s)$, and thus this case does not affect the adiabatic evolution.
(iii) In the limit $\omega \rightarrow 0$, the initial diabatic states $\ket{\Psi_J;0}$ and $\ket{\Psi_J;1}$ become pairwise degenerate.
If $\ket{\Psi_0;0}$ and $\ket{\Psi_0;1}$ are coupled through $\hH(s)$,  then 
the two states will hybridize in the $\omega \rightarrow 0$ limit to form a polaritonic state, 
and the excited adiabatic state preparation will become impossible.
This situation occurs when the photon polarization vector transforms as the 
totally symmetric irreducible representation, but can be avoided using certain definitions of the path
$\omega(s)$ and $\lambda(s)$, as described below.

% PATH DEFINITION
We now consider the form of the parametrized adiabatic pathway.
This pathway must satisfy the boundary conditions $\omega(0)=0$ and $\lambda(0)=\lambda(1)= 0$, meaning that $\lambda(s)$ must be non-linear.
These conditions prevent simply ``switching on'' the electron-photon interaction as $\hH(s) = \hHe + s \hH_\text{int}$, where $\hH_\text{int}$ includes the photon-dependent terms, 
since then the light-matter coupling would not vanish for $s=1$.
Instead, we propose the smooth parametrized adiabatic pathway
\begin{equation}
\omega(s) = \wmax s \quad\text{and}\quad \lambda(s) = \lmax \sin[3](\pi s),
\label{eq:path}
\end{equation}
where $\wmax$ and $\lmax$ are hyperparameters, $s\in[0,1]$, and the boundary condition $\lambda(0)=\lambda(1)=0$ is satisfied.
Here, using $\sin[3](\pi s)$ rather than $\sin(\pi s)$ is required to avoid forming a polaritonic state for $s \rightarrow 0$ by ensuring 
that the non-adiabatic coupling remains zero in this limit (Supplementary Note~1).
The value of $\lmax$ can be varied to control the strength of any avoided crossings (Fig.~\ref{fig:twolevel}\textcolor{blue}{b}; right).
The parameter $\wmax$ should be chosen such that the  avoided crossing between the ground and 
target excited state occurs at approximately $s = \frac{1}{2}$, where $\lambda(s)=\lmax$, 
ensuring the maximal splitting of the avoided crossing and thus minimizing the amount of nonadiabatic transfer.
Typically this requires $\wmax \approx 2 \Delta E$, where $\Delta E$ is the corresponding excitation energy
that can be estimated using quantum subspace expansion or linear response
theory.\cite{McClean2017,Colless2018,Ollitrault2020,Urbanek2020,Asthana2023,Jensen2024,Reinholdt2024,Gandon2024}
However, for sufficiently long time evolution, the final state fidelity only weakly depends on $\wmax$ and $\lmax$ 
and we simply require $\wmax > \Delta E$ to get an accurate final state, as shown for the two-level model in Supplementary Figure 1.
Therefore, the total evolution time provides a tunable parameter to systematically improve the final state fidelity in cases where a good estimate of $\Delta E$ is not available.
Furthermore, if the avoided crossing is not traversed correctly, then the photon number of the final state will be significantly non-zero, providing a metric to 
check the success of adiabatic state preparation when approximate parameters are used.
%The time scale for adiabatic evolution must be long enough given the energy scale of the system.
%As such, we express the evolution time in inverse energy units, adapted to the natural energy scale of each system.

% TIME EVOLUTION
We demonstrate the full excited adiabatic state preparation for the two-level model in 
Eq.~\eqref{eq:PFham} with 
the electronic Hamiltonian parameters $\epsilon=1$ and $\mu=1$, and the path parameters 
$\wmax = 4$, $\lmax = 0.5$ and $N=100$ using a statevector simulation.
Starting from the diabatic state $\ket{0;1}$ with $E=-\epsilon$, 
the energy initially increases linearly as $\omega(s)$ becomes non-zero  (Fig.~\ref{fig:twolevel}\textcolor{blue}{c}; top).
Around $s=\frac{1}{2}$, there is an avoided crossing with the $\ket{1;0}$ diabatic state and 
the energy plateaus at the excited-state energy $E=\epsilon$.
The fidelity $\abs{\braket{1;0}{\Psi(s)}}^2$ confirms the successful preparation of the 
target excited state (Fig.~\ref{fig:twolevel}\textcolor{blue}{c}; bottom).
The accuracy of the adiabatic process improves for longer time evolution, 
as demonstrated by comparing the energy and fidelity for different total times $T=5$, 10, and 50.
Since ASP is expected to be a precursor to fault-tolerant QPE,
we only require a sufficiently high target fidelity, assumed to be $\sim 75\,\%$ in 
Ref.~\onlinecite{Lee2023}.
The two-level system considered here demonstrates how this target fidelity for an 
excited state can be achieved using the EXASP procedure.

%%%%%%%%%%%%%%%%%%%%%%%%
\subsection{Implementation on a quantum device}
%%%%%%%%%%%%%%%%%%%%%%%%

To implement EXASP on real quantum hardware, we first require the 
initial ground state to be prepared on the quantum device.
Established ground state preparation techniques can be used, such as ASP or VQE,
with standard fermion-to-qubit transformations.\cite{Jordan1928,Bravyi2002}
The two-level photon mode is trivially encoded using a single qubit, which is initialized in the 
state $\ket{1}$ to represent the ground state coupled to a photon, i.e.\ $\ket{\Psi(0)} = \ket{\Psi_0}\otimes\ket{1}$.
A second Trotter approximation is required to implement the time-evolution 
steps [Eq.~\eqref{eq:Us}] because $\hH(s)$ generally contains non-commuting terms. 
This additional Trotter approximation may require a smaller time step,
which we investigate in the next section \textit{\textbf{Optical gap for strongly-correlated Hubbard chains}}.

Once the adiabatic time evolution from $s=0$ to 1 is complete, the quantum register 
is expected to be in the final state $\ket{\Psi(1)} = \ket{\Psi_\text{es}} \otimes \ket{0}$, where $\ket{\Psi_\text{es}}$ 
is the target electronic excited state.
However, if the adiabatic evolution is not perfect, then the final state will (to a first approximation) 
correspond to the linear combination 
$\ket{\Psi(1)} \approx a\ket{\Psi_0}\otimes\ket{1} + b \ket{\Psi_\text{es}} \otimes \ket{0}$, 
where the electronic excitation has only partially occurred.
%(There may also be contributions from other excited states.)
The fidelity of the excited state preparation can be enhanced by 
using post-selection to project the final state into the photon vacuum state, 
with the fidelity enhancement depending on the accuracy of the adiabatic state preparation.

\begin{figure*}[htb!]
\includegraphics[width=\linewidth]{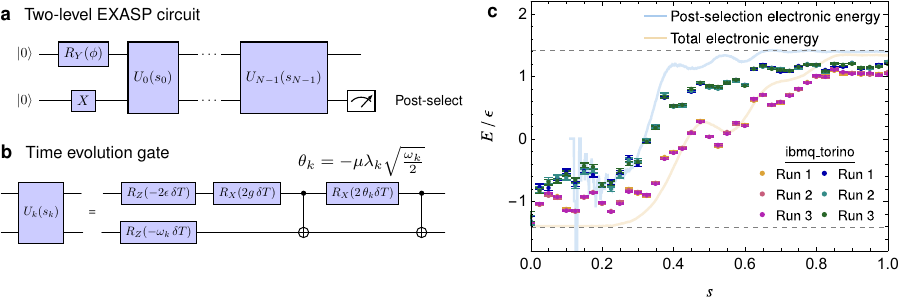}
\caption{\textbf{Implementation of EXASP on quantum hardware for a two-level system.}
(\textbf{a}) Circuit implementation including initial ground state preparation, 
time-evolution gates, and post-selection for the photon vacuum state.
(\textbf{b}) Gate decomposition of the  time-evolution step for the two-level system.
(\textbf{c}) Electronic energy expectation values for quantum hardware experiments using 
\texttt{ibmq\_torino} with $\delta T = 0.5\,\epsilon^{-1}$ (points), compared to a noiseless 
simulation with $\delta T = 0.01\,\epsilon^{-1}$ (solid), showing successful adiabatic preparation
of the excited state.
The system parameters are $\epsilon = 1$, $g=1$, $\mu = 1$, $T=20$, $\wmax=5$, and $\lmax=1$.
}
\label{fig:twolevel_hardware}
\end{figure*}

To investigate the performance of excited adiabatic state preparation, 
we consider a two-level electronic system in the non-eigenstate basis with Hamiltonian coupling $g$ 
between the two basis states. 
The Hamiltonian and dipole operators are defined as
\begin{equation}
\hHe = \begin{pmatrix}
-\epsilon & g \\ g & \epsilon
\end{pmatrix}
\quad\text{and}\quad
\hat{\mu} = \begin{pmatrix}
0 & \mu \\ \mu & 0
\end{pmatrix}.
\end{equation}
We use the non-eigenstate basis to create a more realistic simulation and to avoid the trivial case where the excited state can be prepared using a 
single $X$ gate applied to the ground state, represented by the $\ket{0}$ qubit state.
This two-level system coupled to a single two-level photon mode can be encoded using two qubits, 
with the Hamiltonian defined in terms of qubit Pauli operators as
\begin{equation}
%\begin{split}
\hH(s) = 
-\epsilon\, Z I + g XI + \frac{\omega(s)}{2} (II - IZ) 
%\\
- \lambda(s)\, \mu \sqrt{\frac{\omega(s)}{2}} XX + \frac{\lambda(s)^2 \mu^2}{2} II.
%\end{split}
\label{eq:twolevel_qubit}
\end{equation}
%where again the dipole-self energy contribution ($\frac{1}{2} \lambda^2 \mu^2 II$) provides a 
%uniform energy shift
The ground electronic state can be encoded using an $R_Y(\phi)$ rotation on the first qubit with 
$\phi = \tan[-1](-g/\epsilon)$, while the photon qubit is initialized in the $\ket{1}$ state
(Fig.~\ref{fig:twolevel_hardware}\textcolor{blue}{a}).
The time-evolution operator for step $k$ is encoded with the Trotter product formula
\begin{equation}
\hat{U}_k(s_k) 
\approx
\mathrm{e}^{\mathrm{i}\mu \lambda_k \sqrt{\frac{\omega_k}{2}}\, \delta T XX}
\mathrm{e}^{\mathrm{i}\frac{\omega_k}{2}\delta T IZ}
\mathrm{e}^{-\mathrm{i} g \delta T XI}
\mathrm{e}^{\mathrm{i} \epsilon \delta T ZI},
\end{equation}
where $\omega_k = \omega(s_k)$ and $\lambda_k = \lambda(s_k)$.
Constant terms in $\hH(s)$
provide only a global phase-shift and can be ignored from the time evolution.
The corresponding circuit implementation is shown in Fig.~\ref{fig:twolevel_hardware}\textcolor{blue}{b}.

%%%%%%%%%%%%%%%%%%%%%%%%%%%%%
%%%%%%%%%%%%%%%%%%%%%%%%%%%%%

\begin{figure}[htb!]
\includegraphics[width=0.5\linewidth]{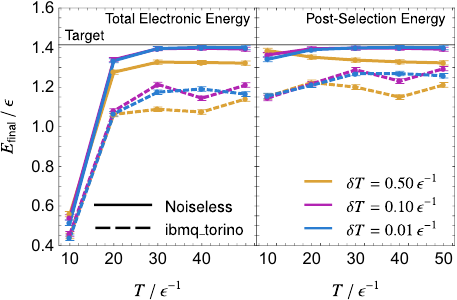}
\caption{\textbf{Convergence of the two-level EXASP final electronic energy with and without post-selection.}
Comparison of noiseless statevector simulations with hardware calculations on the \texttt{ibmq\_torino} quantum chip.
Parameters used were $\epsilon=1$, $g=1$, $\wmax = 5$, $\lmax = 1.0$, and $\mu=1$ with $10^5$ measurement shots. 
}
\label{fig:twolevel_errors}
\end{figure}

Simulations on near-term hardware are restricted to shallow quantum circuits, meaning
that only a few adiabatic steps can be achieved.
We have computed the total electronic energy (without photonic contributions) and
the post-selection electronic energy along the adiabatic evolution pathway using 
the \texttt{ibmq\_torino} quantum chip,
with a total evolution time $T=20\,\epsilon^{-1}$ and $\delta T = 0.5\,\epsilon^{-1}$ 
(Fig.~\ref{fig:twolevel_hardware}\textcolor{blue}{c}).
These hardware results demonstrate successful adiabatic evolution from the 
ground state to the excited state.
The time step error is evident through comparison to a noiseless simulation 
with $\delta T=0.01\,\epsilon^{-1}$.
Initially, the post-selected electronic energy represents noise in the photon qubit state, 
with the electronic energy expectation value centred close to the ground-state energy.
However, once the adiabatic evolution starts to evolve into 
the $\ket{\Psi_\text{es}}\otimes\ket{0}$ state, 
the post-selected energy converges towards the excited-state energy more rapidly
than the total (unprojected) electronic energy. 
This faster convergence demonstrates the fidelity enhancement achieved by 
projecting into the zero photon state, 
%, providing a straightforward route to improve the 
% success probability of the excited state preparation.

The non-zero time step, finite evolution time, and number of measurement shots all affect the
uncertainty or accuracy of the EXASP procedure.
Noisy simulations for the \texttt{ibmq\_torino}  quantum chip show that the measurement uncertainty 
in the two-level system is converged with $10^5$ shots (Supplementary Figure 2).
The final electronic energy estimate converges rapidly with both the total evolution time and 
the time step, while the photon post-selection significantly improves the accuracy for short 
time evolution
(Fig.~\ref{fig:twolevel_errors}).
Calculations on the \texttt{ibmq\_torino} quantum chip show that current hardware noise reduces 
the accuracy of the final electronic energy estimate (Fig.~\ref{fig:twolevel_errors}). 
However, the adiabatic time evolution for the two-level system is simple enough that the 
circuit compilation reduces the excited adiabatic state preparation to only two 2-qubit gates for any $T$ and $\delta T$
(Supplementary Figure~3), meaning that the effect of noise does not grow with the number of time steps in this case.

%%%%%%%%%%%%%%%%%%%%%%%%%%%%%
%%%%%%%%%%%%%%%%%%%%%%%%%%%%%

%%%%%%%%%%%%%%%%%%%%%%%%
\subsection{Optical gap for strongly-correlated Hubbard chains}
%%%%%%%%%%%%%%%%%%%%%%%%

The EXASP procedure can also be applied to realistic Hamiltonians,
 such as the Hubbard model
of strongly-correlated solid-state physics.
Here, we consider a one-dimensional Hubbard chain with the electronic Hamiltonian
\begin{equation}
\hat{H}_\text{e} = - t\sum_{\langle p,q \rangle} (\hat{a}_{p\uparrow}^\dagger \hat{a}_{q\uparrow} + \hat{a}_{p\downarrow}^\dagger \hat{a}_{q\downarrow}) + U \sum_p \hat{n}_{p\uparrow} \hat{n}_{p\downarrow},
\label{eq:hub_ham}
\end{equation}
where $\hat{a}^\dagger_{p\sigma}$  ($\hat{a}_{p\sigma}$) are the creation (annihilation) operators 
for electrons with spin $\sigma$ at lattice site $p$, 
the number operators are $\hat{n}_{p\sigma} = \hat{a}^\dagger_{p\sigma}\hat{a}_{p\sigma}$,
and the double sum $\sum_{\langle p,q \rangle}$ is restricted to nearest-neighbour sites (i.e., $q=p\pm1$) 
with open boundary conditions.
The ratio of the on-site electron repulsion $U$ and the one-electron hopping $t$ determines the 
electron correlation strength.
An $N$-site Hubbard chain can be encoded using $2N$ qubits using the Jordan--Wigner encoding,\cite{Jordan1928} 
with an additional qubit  encoding the photon state necessary for EXASP. 
We consider the 4- and 6-site Hubbard chains as they are sufficiently complex to 
explore the effects of the time step, Trotterization error, dark states, and the use of
inexact starting wavefunctions on the EXASP propagation. 
We consider a range of correlation strengths between $U=1t$ (weak correlation) 
and $U=8t$ (strong correlation). 
Since these systems can be solved exactly, we always set $\omega_\text{max} = 2 \Delta E$ using the excitation
energy $\Delta E$ for the target state, and $\lambda_\text{max} = 1$.
These parameters are found to be sufficient to give satisfactory excited state preparation, although they are not necessarily optimal.

\begin{figure}[htb!]
\includegraphics[width=0.6\linewidth]{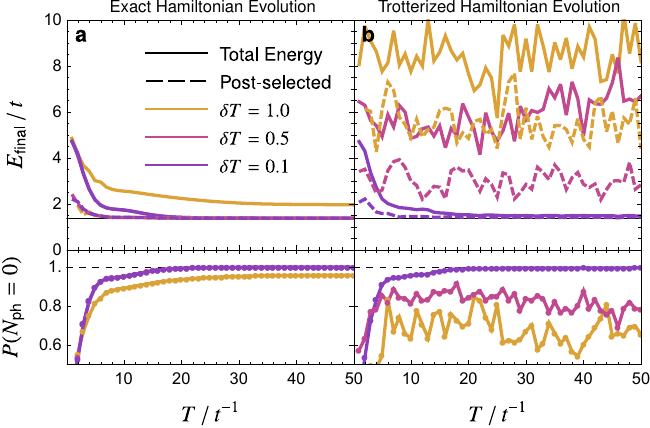}
\caption{
\textbf{Convergence of the final energy and success probability of projecting into the
$\ket{0}$ photon state for the 4-site Hubbard chain with $U=4t$.}
(\textbf{a}) Final energy after exact time evolution for different time steps with $U = 4t$, with and without photon state post-selection.
(\textbf{b}) Approximating the propagation using a Trotter expansion of the time-evolution operator requires a smaller time step to obtain satisfactory convergence.
}
\label{fig:hubbard-4site}
\end{figure}

Finding a suitable time step that is large enough to avoid long propagation 
times, but not so large that it introduces singificant Trotterization error, is critical 
to adiabatic state preparation.
We use the 4-site Hubbard model to investigate how the time step affects the 
final accuracy of the EXASP evolution. 
For a moderately correlated lattice ($U = 4t$) initialized with the exact ground 
state wavefunction, we find systematic convergence in the final state energy (including photon contributions)
as the total time $T$ increases or the time step $\delta T$ decreases (Fig.~\ref{fig:hubbard-4site}\textcolor{blue}{a}). 
The convergence with respect to $T$ is accelerated using a smaller time step, 
giving more than 96\% fidelity with the target state obtained
for $T=10\,t^{-1}$ and $\delta T \leq 0.5\,t^{-1}$
when projecting onto the $\ket{0}$ photon state before measuring the energy. 
As expected, the accuracy of the final energy is consistently enhanced by
this final state projection, 
with over 90\% probability of successfully selecting the $\ket{0}$ photon
state at $T=10\,t^{-1}$ even though the total 
energy does not converge until $T \approx 20\,t^{-1}$.
Therefore, post-selection on the photon vacuum state
successfully accelerates the convergence of the EXASP propagation with respect 
to the total evolution time for realistic Hamiltonians.

\begin{figure*}[htb!]
 \includegraphics[width=\linewidth]{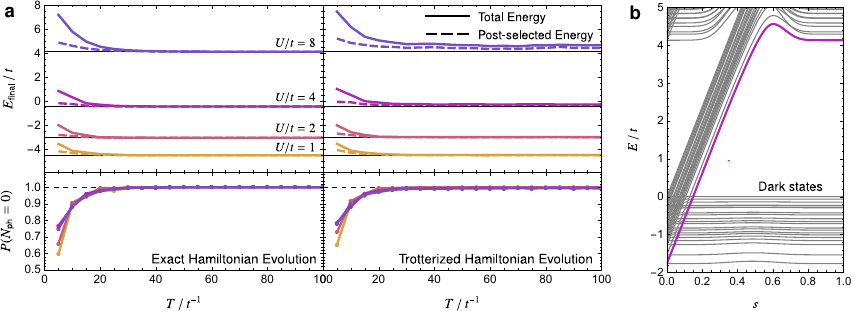}
\caption{\textbf{EXASP propagation for the 6-site Hubbard chain. }
(\textbf{a}) Convergence of the final total energy and probability of measuring the $\ket{0}$ photon state at various $U/t$ values as a function of total evolution time $T$ with $\delta T = 0.1\,t^{-1}$. Similar convergence is obtained if the propagation is performed exactly or using a first-order Trotter expansion.
(\textbf{b}) Adiabatic evolution along the EXASP pathway (purple) for $U/t = 8$, $T = 100\,t^{-1}$ and $\delta T = 0.1\,t^{-1}$, shown alongside the eigenvalues of the full system Hamiltonian with the dipole self-energy included (grey).  EXASP leads to the first bright excited state as there is no coupling to the lower energy dark states.
}
\label{fig:hubbard-6site}
\end{figure*}

Trotterization of the time-evolution operator imposes stronger conditions 
on the  time step to avoid excessive errors within each time-evolution step. 
For all the Hubbard systems considered here, $\delta T = 0.1\,t^{-1}$ is sufficiently 
small to allow nearly exact Trotter propagation (Fig.~\ref{fig:hubbard-4site}\textcolor{blue}{b}). 
The observed trends for the exact Hamiltonian simulation continue to hold in 
this case, with convergence of the post-selected energy requiring approximately 100 applications of the time-evolution operator.  
In contrast, the larger time steps considered ($\delta T = 0.5\,t^{-1}$ and $1.0\,t^{-1}$) lead to a final 
energy that does not converge in the long $T$ limit.
Notably, the time step $\delta T = 0.1\,t^{-1}$ is also sufficient to accurately propagate the 4-site Hubbard model across the entire range of correlation strengths
considered (Supplementary Figure~4), as well as the larger 6-site 
Hubbard system (Fig.~\ref{fig:hubbard-6site}\textcolor{blue}{a}).
We expect that smaller time steps may be required for more complex Hamiltonians
due to the presence of more non-commuting terms.

Next, we consider how the EXASP procedure changes for larger systems using 
the 6-site Hubbard model.
As the size of the system increases, the total time required to accurately prepare an excited state also increases. 
For example, in the $U = 4t$ case, $T \approx 25\,t^{-1}$ is necessary for a final state fidelity greater than 96\% after
post-selecting the $\ket{0}$ photon state for the 6-site model, compared to $T=10\,t^{-1}$ for 4 sites.
We find that propagation for longer total time is required 
when the electron correlation $U/t$ is stronger (Fig.~\ref{fig:hubbard-6site}\textcolor{blue}{a}; left), 
%Convergence is also dependent on the degree of correlation in the system, with larger $U/t$ values requiring longer propagation
%times (Fig.~\ref{fig:hubbard-6site}\textcolor{blue}{b} left). 
and that the Trotterization error also increases with $U/t$ (Fig.~\ref{fig:hubbard-6site}\textcolor{blue}{a}; right).
For example, the final state energy at $U/t = 8$ has an error of $13\%$ that persists  at large $T$, even after post-selection into the $\ket{0}$ photon state,
%Together with the convergence of the success probability of selecting the 
%photonless state to unity, this suggests 
suggesting that Trotterization of the time-evolution step can introduce contamination
from other eigenstates that are also coupled to the $\ket{0}$ photon state. %of the underlying system, even as it correctly annihilates the photon. 

The larger 6-site system also provides an example of how the EXASP propagation
can pass over low-energy dark states to reach the first bright state, despite the presence
of exact degeneracies along the adiabatic pathway.
In the strongly correlated Hubbard chain, the low-lying excited states correspond to 
different spin arrangements where every site is singly occupied.
These states have negligible dipole coupling to the antiferromagnetic ground state, 
meaning that
they do not affect the adiabatic propagation due to the lack of nonadiabatic coupling.
Instead, adiabatic state preparation with exact time propagation evolves from the ground state to the lowest-energy bright state, 
which includes a doubly-occupied site (Fig.~\ref{fig:hubbard-6site}\textcolor{blue}{b}).
Therefore, the EXASP approach naturally computes the excited-state wavefunction and energy 
associated with the optical gap, which is a key quantity for understanding light-matter 
interactions in optoelectonics and photovoltaics.

\begin{figure}[htb]
\includegraphics[width=0.5\linewidth]{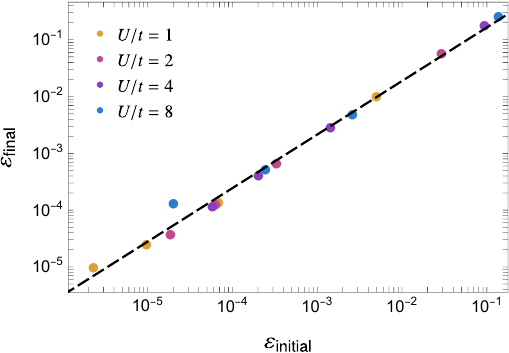}
\caption{\textbf{Correlation between the initial and final state fidelity error for
 EXASP in the 6-site Hubbard chain.}  States were
	propagated for $T = 100\,t^{-1}$ with $\delta T = 0.1\,t^{-1}$. For each value of
	$U/t$, the different points correspond to a pp-tUPS ground state obtained with 1--4
	layers, and the error decreases monotonically as the number of layers increases.}
\label{fig:hubbard_6site_tups}
\end{figure}

To initialize the EXASP procedure in practice, we rely on an approximate ground state 
prepared on the quantum device using a method such as VQE or ASP. 
To assess how the accuracy of this initial ground state affects the fidelity of the 
final excited state, we initialize the system using ground-state
approximations with varying quality.
In this work, we consider approximate ground states obtained through VQE as these have 
have known quantum circuit implementations and their errors have been previously characterised for the Hubbard chain\cite{Burton2024b}.
We use the tiled unitary product state (tUPS) \textit{ansatz}\cite{Burton2024} 
for which the number of \textit{ansatz} layers provides systematic convergence to the exact ground state.
In all cases, we consider the perfect pairing tUPS (pp-tUPS) variant
(see \textit{Methods} section), which is known to converge rapidly with 
respect to the number of layers for the Hubbard model.\cite{Burton2024b}
As expected, increasing the number of \textit{ansatz} layers leads to systematically
higher initial state fidelity  (Supplementary Figure~5).
Furthermore, for sufficiently long EXASP propagation times, 
the initial state fidelity error  $\varepsilon_{\text{initial}} = 1 - \abs{\braket{\Psi_0}{\Psi_\text{initial}}}^2$ is  proportional to the fidelity error 
of the final state $\varepsilon_{\text{final}}= 1 - \abs{\braket{\Psi_J}{\Psi_\text{final}}}^2$
with the target excited state $\ket{\Psi_J}$.
(Fig.~\ref{fig:hubbard_6site_tups}). While the quality of a given tUPS 
approximation decreases as the correlation in the system increases, the
relationship between the initial and final state fidelity error is independent of the correlation
strength and can be fitted to a near-linear form $\varepsilon_\mathrm{final} \approx 1.448 \varepsilon_\mathrm{initial}^{0.943}$.
We note that in the long-time limit, post-selection on the photon mode has little effect on the fidelity with the
target excited state, so this observed error propagation can be attributed almost exclusively to the initial
state error, rather than incomplete transfer between the $\ket{1}$ and $\ket{0}$ photon states.

We have attempted a hardware implementation of EXASP for the two-site
Hubbard model to move beyond the two-level system tested previously.
This system can be mapped to three qubits if only the 
two-electron $m_s = 0$ subspace is considered (Supplementary Note~2).
While we were able to run these calculations on the \texttt{ibmq\_torino}  quantum 
chip, the level of device noise was too high to measure any significant 
preparation of the excited state  (Supplementary Figure~6).
This result demonstrates that current hardware is still too noisy
for methods  that move beyond the near-term intermediate scaling regime,
such as adiabatic state preparation.

%%%%%%%%%%%%%%%%%%%%%%%%
\subsection{Molecular excitations using symmetry selection rules}
%%%%%%%%%%%%%%%%%%%%%%%%

%%%%%%%%%%%%%%%%%%%%%%%%

\begin{figure*}[htb!]
\includegraphics[width=\linewidth]{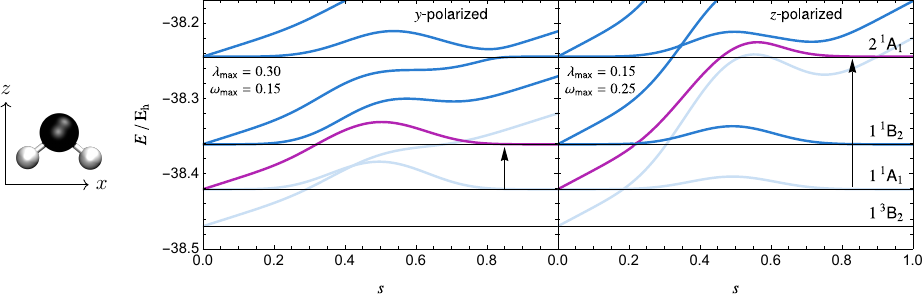}
\caption{\textbf{EXASP can target different excited states by varying the photon polarization vector.}
Exact eigenstates of the coupled electron-photon Hamiltonian for \ce{CH2} with a single photon mode 
polarized aligned along the $y$-direction (left) 
or $z$-direction (right).
The target state (purple) evolves from the singlet ground state to the $1\,^1\text{B}_2$ or $2\,^1\text{A}_1$ states using the $y$- or $z$-polarization, respectively.
The triplet ground state (light blue) does not couple to the singlet states and can be ignored.}
\label{fig:figure7}
\end{figure*}

\begin{figure}[htb]
\includegraphics[width=0.65\linewidth]{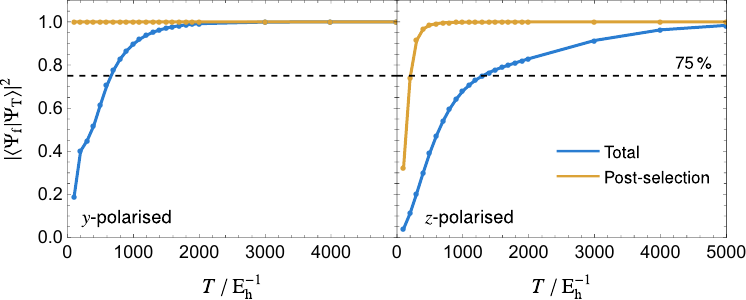}
\caption{\textbf{Fidelity of the excited state preparation for \ce{CH2} systematically converges with increasing total time.}
The final state fidelity is shown for $\delta T=1.0\,\mathrm{E_h}^{-1}$ 
with the photon polarization aligned along the $y$-direction ($\lmax = 0.30$, $\wmax = 0.15$) or the $z$-direction ($\lmax = 0.15$, $\wmax = 0.25$). 
Post-selection for the $\ket{0}$ photon state significantly increases the fidelity.}
\label{fig:figure8}
\end{figure}

In molecules with non-trivial point group spatial symmetry, it is often desirable to  
prepare excited states that transform as different irreducible representations.
Targeting particular state symmetry is achieved using EXASP by changing the 
photon polarization vector and exploiting the dipole coupling symmetry selection rules.
This approach can be illustrated for the low-lying $\mathrm{1\,^1A_1}$, $\mathrm{1\,^1B_2}$, and $\mathrm{2\,^1A_1}$ singlet states  in methylene,
where the molecule is oriented in the $xz$-plane with the principal rotation axis in the $z$ direction.
Starting the propagation in the singlet ground state,  the 
$\mathrm{1\,^1B_2}$ or $\mathrm{2\,^1A_1}$ excited states can be systematically 
prepared by orienting the photon polarization vector along the $y$ or $z$ directions, respectively, as illustrated by the parametrized
eigenstates along suitable adiabatic pathways in Fig.~\ref{fig:figure7}.
Here, we have chosen $\lmax$ and $\wmax$ by using trial and error to maximize the final state fidelity.
In a realistic quantum setting, it would be possible to optimize $\lmax$ and $\wmax$ by minimizing the 
photon number in the final state, which is independently measurable and 
negatively correlated with the final state fidelity (Supplementary Figure 7).
We have numerically computed the excited-state fidelity using exact 
time evolution  (Fig.~\ref{fig:figure8}), confirming that the adiabatic preparation of 
particular symmetry states can be achieved in practice.

%%%%%%%%%%%%%%%%%%%%%%%%

Methylene also provides a molecular example to investigate how the final excited-state 
fidelity depends on the total time and the time step for the adiabatic evolution. 
The time evolution was found to be converged for a time step of 
$\delta T = 1.0\,\mathrm{E_h}^{-1}$.
Due to the larger Hilbert space and number of Hamiltonian terms in methylene,
the effect of using a Trotter approximation for each propagation step could 
not be feasibly tested. 
For the $y$- and $z$-polarized cases, total evolution times of $T=700\,\mathrm{E_h}^{-1}$ and $T=1300\,\mathrm{E_h}^{-1}$ are necessary to
reach a target fidelity of $75\,\%$ (Fig.~\ref{fig:figure8}), 
which is considered to be sufficient to use ASP as a precursor to QPE.\cite{Lee2023}
Projection into the $\ket{0}$ photon state using post-selection significantly increases
this fidelity. 
However, this increase in fidelity is entirely offset by the probability of measuring the
$\ket{0}$ photon state, meaning that the overall probability of preparing the target excited
state after QPE is the same regardless of whether photon projection is used.
Nevertheless, this post-processing of the EXASP final state provides a practical route 
to screen for high-fidelity states. 
This screening allows the algorithm to fail early if
post-selection fails, enhancing the success
%creating a `fail early' protocol to enhance the success
probability of a subsequent excited-state QPE calculation and minimizing the amount of wasted quantum resources.
%{\andreea{MAF: We could comment here that this allows for a `fail early' protocol, which
%saves quantum resources. Also if the excited state energy is unknown, this gives us more
%knowledge of whether the QPE has selected the correct eigenstate.}}
These results confirm that EXASP can be applied to selectively prepare 
the lowest-energy molecular excited state for different irreducible representations, 
which is key for applications to photochemistry.

%%%%%%%%%%%%%%%%%%%%%%%%
%%%%%%%%%%%%%%%%%%%%%%%%

%%%%%%%%%%%%%%%%%%%%%%%%
\section{Discussion}
\label{sec:discussion}
%%%%%%%%%%%%%%%%%%%%%%%%

We have shown that excited states can be encoded on a 
quantum device by combining adiabatic state preparation with explicit light-matter coupling.
Our EXASP algorithm prepares the lowest-energy bright state from the
ground-state wavefunction by adiabatically turning on the electron-photon interaction
while increasing the photon frequency.
The excited-state fidelity can be further enhanced by projecting the final state into 
the zero-photon space using post-measurement selection on the photon qubit,
providing an early failure protocol for subsequent quantum phase estimation.
%`fail early' protocol for subsequent quantum phase estimation. 
Furthermore, excited states with different symmetries can be systematically targeted by changing 
the polarization direction of the photon mode.
Numerical simulations demonstrate that high-fidelity excited states can be achieved for
the strongly correlated Hubbard model and the methylene molecule, while experiments
on current quantum hardware illustrate how the method can be implemented in practice.

EXASP offers key advantages for excited state preparation:
it requires minimal information about the target excited-state wavefunction beyond 
an estimate of the excitation energy;
it does not rely on any variational \textit{ansatz} or modified excited-state objective function; 
and it avoids hybrid subspace diagonalization, which can only estimate excitation energies and 
would require a linear combination of unitaries  to prepare an excited state on the quantum device.
%{\andreea{MAF: Only if you want to use it for state preparation, rather than energy estimation. If
%I understand your point correctly.}} 
Furthermore, our algorithm systematically converges to the exact excited state by increasing
the total time evolution and the initial state fidelity, and decreasing the time step.
Increasing the total time evolution also reduces the sensitivity of the algorithm to 
the hyperparameters $\wmax$ and $\lmax$.
This systematic convergence provides a non-heuristic technique to 
prepare high-fidelity excited states for subsequent fault-tolerant algorithms, 
such as QPE or quantum dynamics.
These advances lay the foundation to extend high-accuracy quantum algorithms to 
 challenging excited states in weakly and strongly correlated electronic systems.

We have performed simulations on IBM quantum hardware to test the implementation of 
EXASP  for model Hamiltonians.
While an accurate excited-state energy can be achieved for a two-state model, the
amount of hardware noise on current devices prevents any meaningful results from 
being obtained for more complex Hamiltonians, such as the Hubbard model.
As an adiabatic evolution method, the practicality of EXASP will benefit from further 
advances in fault-tolerant Hamiltonian evolution and adiabatic state preparation, as well
as future hardware developments.

So far we have considered a digital quantum computing implementation of the EXASP algorithm, which leads to
certain  challenges such as the Trotterization error in the time evolution. 
It may be possible to avoid this issue using an analog quantum implementation.\cite{Das2008,Kendon2010} 
For example, using trapped ion architectures, one could directly represent the auxiliary photon using the device bosonic cavity modes instead of a qubit representation. 
In practice, this would require both tunable resonator frequencies\cite{Palacios-Laloy2008,Whittaker2014} and tunable coupling to the qubits\cite{Lu2017} 
to effectively implement the adiabatic pathway. While each of these is possible in isolation, the combined variation of 
both parameters may prove to be a significant engineering challenge.

In the current work, we have only considered the dipole of electronic states to a single photon, 
which can only target the lowest-energy bright state for each optically active irreducible representation.
However, the framework developed here could be readily extended to other types of coupling
that enable different excited states to be accessed.
For example, dipole-forbidden states could be prepared using an artificially engineered strong quadrupole coupling.
Different spin states could be accessed either by starting from the corresponding spin ground state, 
or by introducing a spin-orbit interaction.
Furthermore, extending the photon mode to four energy levels and starting the adiabatic preparation in the 
$\ket{2}$ photon state would allow optically dark double excitations to be prepared \textit{via} an intermediate bright state, 
mirroring a nonlinear two-photon absorption processes.
An alternative approach to access higher energy levels would be to reset the photon qubit to the $\ket{1}$ state after post selection
and repeat the adiabatic process, creating a ladder of excited states.
We leave these extensions for future work.

%Furthermore, we have only considered the coupling of an electronic state to a 
%single photon, which can only target the lowest-energy bright 
%state for each optically active irreducible representation.
%Dark excited states with no dipole coupling to the ground state may be accessed using 
%two or more photons, or by resetting the photon qubit to the $\ket{1}$ state at the end of 
%the propagation and repeating the adiabatic process. 
%We leave these extensions for future work.
%{\andreea{MAF:Also by resetting the qubit to the 1 state after post selection, we can
%construct an excited state ladder.}}

The success of quantum algorithms for photophysics and 
photochemistry relies on the preparation of high-fidelity excited states, 
which is difficult to achieve with VQE or subspace expansions.
Here we have shown how the lowest bright excited state can be adiabatically prepared starting
from the ground state by introducing explicit light-matter coupling. The resulting
EXASP methodology provides a systematically convergent approach to prepare accurate
excited state wavefunctions.
%The EXASP algorithm is systematically convergent to the exact excited state and is
%free from any variational ansatz or subspace expansion.
%{\andreea{MAF: some repetition here with the second paragraph.}}
Our results demonstrate that adiabatic state preparation can be extended to excited electronic
states, laying the foundation for future quantum algorithms to simulate light-matter interactions in 
molecular and materials science.

%%%%%%%%%%%%%%%%%%%%%%%%
\section{Methods}
\label{sec:methods}
%%%%%%%%%%%%%%%%%%%%%%%%
%%%%%%%%%%%%%%%%%%%%%%%%
\subsection{Statevector and noisy simulations}
%%%%%%%%%%%%%%%%%%%%%%%%
All classical simulations were performed using an in-house code 
implemented in python.\cite{Exasp}
For the two-level model, the noiseless and noisy simulations were implemented 
using an interface to the \textsc{Qiskit}\cite{JavadiAbhari2024} statevector simulator.
Exact statevector simulations without explicit circuit sampling were performed for 
the 4- and 6-site Hubbard model with and without Trotterization, and for the 
methylene molecule.
The Hamiltonian and dipole operators for the Hubbard model were defined in the site 
basis.
For the methylene molecule, the Hamiltonian and dipole matrix elements were defined
in the Hartree--Fock molecular orbital basis using the 
STO-3G basis set,\cite{Hehre1969} computed using the \textsc{Quantel} package.\cite{Quantel}
Noisy simulations were performed using the \texttt{AerSimulator} in \textsc{Qiskit} with 
the noise model generated for the \ibmtorino{} quantum chip.

%%%%%%%%%%%%%%%%%%%%%%%%
\subsection{Initial ground state preparation}
%%%%%%%%%%%%%%%%%%%%%%%%
 EXASP simulations were performed starting from the exact or an approximate ground state.
The exact ground state for the two-level system was prepared using a single $R_Y(\phi)$
gate acting on the electronic qubit, with $\phi = \tan^{-1}(-g/\epsilon)$. 
This approach was used in both the simulated calculations and hardware implementation.
For the Hubbard model and the methylene molecule, only the exact (singlet) ground state 
was used  in statevector simulations and the circuit required to 
implement the ground state was ignored.
The approximate ground states considered for the Hubbard model were defined
using the variational perfect-paired tiled unitary product state (pp-tUPS) \textit{ansatz},
which includes orbital optimization,\cite{Burton2024} with an increasing number of layers to control the accuracy. 
In this formalism, the wavefunction is defined for a reference state $\ket{\Phi_0}$ as
\begin{equation}
\ket{\Psi_\text{tUPS}} = 
\underbrace{\prod_{m=1}^{\lceil{N/2}\rceil}
\qty(
\prod_{p=1}^{\lfloor{(N-1)/2}\rfloor} \hat{Q}_{2p+1,2p}^{(m)}
\prod_{p=1}^{\lfloor{N/2}\rfloor}       \hat{Q}_{2p, 2p-1}^{(m)}
)}_{\text{orbital optimization}}
\prod_{m=1}^L 
\qty(
\prod_{p=1}^{\lfloor{(N-1)/2}\rfloor} \hat U_{2p+1, 2p}^{(m)} 
\prod_{p=1}^{\lfloor{N/2}\rfloor}\hat U_{2p, 2p-1}^{(m)}
)\ket{\Phi_0},
\end{equation}
where $L$ is the number of layers and $N$ is the number of spatial orbitals. 
The parametrized unitaries $\hat U_{pq}^{(m)}$ and $\hat Q_{pq}^{(m)}$ are given by
\begin{subequations}
\begin{align}
\hat U_{pq}^{(m)} &= 
\exp(\theta_{pq,1}^{(m)}\hat \kappa_{pq}^{(1)}) 
\exp(\theta_{pq,2}^{(m)}\hat \kappa_{pq}^{(2)})
\exp(\theta_{pq,3}^{(m)}\hat \kappa_{pq}^{(1)}),
\\
\hat Q_{pq}^{(m)} &= \exp(\theta_{pq,1}^{(m)}\hat \kappa_{pq}^{(1)}),
\end{align}
\end{subequations}
where $\kappa_{pq}^{(1)} = \hat E_{pq} - \hat E_{qp}$ and $\kappa_{pq}^{(2)} = \hat E_{pq}^2 - \hat E_{qp}^2$. 
Here, \mbox{$\hat E_{pq} = \hat p^\dagger \hat q + \hat{\bar p}^\dagger\hat{\bar q}$} is the singlet excitation operator from orbital $q$ to $p$ and $E_{pq}^2$ is a paired double excitation from $q$ to $p$. 
In the perfect-pairing variant of tUPS, the initial qubit register alternates between occupied and unoccupied orbitals 
to maximize the correlation captured by a shallow circuit, as described in Ref.~\citenum{Burton2024}.
For our Hubbard simulations, the initial orbitals  were ordered sequentially in the chain as $(\phi_1, \phi_2, \dots, \phi_N)$,
where $\phi_i$ is the spatial orbital on site $i$, as this structure allows the tUPS circuit to exploit the locality of electronic interactions.
The reference state $\ket{\Phi_0}$ for the pp-tUPS \textit{ansatz} then corresponds to the Slater determinant
$\qty|\phi_1 \bar{\phi}_1 \phi_3 \bar{\phi}_3 \cdots |$, where the overbar denotes a low-spin orbital,
although including orbital optimization reduces the dependency on the initial occupied orbitals.
The optimized tUPS wavefunctions were converted to a statevector representation, which was used as the input for EXASP 
in place of the exact ground state.

%%%%%%%%%%%%%%%%%%%%%%%%
\subsection{Computation of tiled unitary product states}
%%%%%%%%%%%%%%%%%%%%%%%%
Following the method described in Ref.~\citenum{Burton2024}, 
the optimal parameters for the tUPS circuit were obtained through statevector simulations
using  global optimization with the basin hopping parallel tempering (BHPT) 
algorithm,\cite{StrodelLWW10,Li1987,Wales1997}
as implemented in the \textsc{GMIN} software.\cite{gmin}
This BHPT scheme used eight replica basin-hopping calculations with temperatures distributed exponentially 
between $0.0001\,t$ and $0.01\,t$, 
and exchange between replicas was attempted with a mean frequency of ten steps. 
Here, $t$ is the magnitude of the Hubbard model hopping term in \Cref{eq:hub_ham} and is used as the natural unit of energy.
On each basin-hopping step, the L-BFGS
optimization algorithm\cite{Broyden1970,Fletcher1970,Goldfarb1970,Shanno1970} 
was performed using analytic gradients, 
with the convergence criteria set to a root-mean-square gradient value below $10^{-5}\,t$ and a maximum of 2000 iterations.
A total of 250 basin hopping steps were performed for each replica and 
the lowest energy minimum was used to define the initial ground state.

%%%%%%%%%%%%%%%%%%%%%%%%
\subsection{Quantum hardware implementation}
%%%%%%%%%%%%%%%%%%%%%%%%
Hardware calculations for the two-level excited state preparation were performed 
using the \ibmtorino{} quantum chip, 
which is one of the IBM Quantum Heron processors.
The circuit was obtained using the transpilation and optimization routines in \textsc{Qiskit}.
Since the experiment only requires two qubits, the circuit was duplicated 50 times 
in parallel to 
maximize the use of this 133 qubit chip and maximize the number of affordable 
measurement shots.
200 measurement shots were performed, giving a total of $10^5$ measurements using
the circuit parallelization.

%\subsection{notes}
%For methylene, $\wmax$ and $\lmax$ have been chosen empirically such that the avoided crossing between the initial and target states occurs
%close to $s = 1/2$ and the nonadiabatic coupling is not too large for adiabatic state %preparation.

%%%%%%%%%%%%%%%%%%%%%%
\section*{Data availability}
%%%%%%%%%%%%%%%%%%%%%%%%

The data required to reproduce figures are hosted in a publicly available repository
[\href{https://doi.org/10.5281/zenodo.17666190}{10.5281/zenodo.17666190}].
All other data required are present in the paper or in the Supplementary Material.

%%%%%%%%%%%%%%%%%%%%%%
\section*{Code availability}
%%%%%%%%%%%%%%%%%%%%%%%%

The code used to perform simulations is hosted in a public GitHub repository at \href{https://github.com/hgaburton/exasp.git}{https://github.com/hgaburton/exasp.git}.

%%%%%%%%%%%%%%%%%%%%%%
\section*{References}
\bibliography{manuscript}
%%%%%%%%%%%%%%%%%%%%%%

%%%%%%%%%%%%%%%%%%%%%%
\section*{Acknowledgements}
%%%%%%%%%%%%%%%%%%%%%%%%

H.G.A.B.\ was supported by Downing College, Cambridge through the Kim and 
Julianna Silverman Research Fellowship and is a Royal Society University Research Fellow (URF\textbackslash{}R0\textbackslash{}241299) at University College London.
M.A.F.\ acknowledges financial support from Peterhouse College, Cambridge through a Research Fellowship.
We acknowledge the use of IBM Quantum services for this work. 
The views expressed are those of the authors, and do not reflect the official policy or position of IBM or the IBM Quantum team.

%%%%%%%%%%%%%%%%%%%%%%%%
\section*{Author contributions}
%%%%%%%%%%%%%%%%%%%%%%%%
H.G.A.B.\ conceived the project and managed the collaboration.
H.G.A.B.\ and M.A.F.\ jointly designed and implemented the code, carried out numerical simulations, 
performed the data analysis, and contributed to preparing and reviewing the manuscript.

%%%%%%%%%%%%%%%%%%%%%%%%
\section*{Competing interests}
%%%%%%%%%%%%%%%%%%%%%%%%

The authors declare no competing interests.

\end{document}